# Caching or No Caching in Dense HetNets?

M. G. Khoshkholgh*, Keivan Navaie**, Kang G. Shin††, V. C. M. Leung*, and Halim Yanikomeroglu†
* Department of Electrical and Computer Engineering, the University of British Columbia, Canada
** School of Computation and Communications, Lancaster University, UK
† Department of System and Computer Engineering, Carleton University, Canada
†† Electrical Engineering and Computer Science Department, University of Michigan, USA

*Abstract*— **Caching the content closer to the user equipments (UEs) in heterogenous cellular networks (HetNets) improves user-perceived Quality-of-Service (QoS) while lowering the operators backhaul usage/costs. Nevertheless, under the current networking strategy that promotes aggressive densification, it is unclear whether cache-enabled HetNets preserve the claimed cost-effectiveness and the potential benefits. This is due to 1) the collective cost of caching which may inevitably exceed the expensive cost of backhaul in a dense HetNet, and 2) the excessive interference which affects the signal reception irrespective of content placement. We analyze these significant, yet overlooked, issues, showing that while densification reduces backhaul load and increases spectral efficiency in cache-enabled dense networks, it simultaneously reduces cache-hit probability and increases the network cost. We then introduce a caching efficiency metric, area spectral efficiency per unit spent cost, and find it enough to cache only about $3\%$ of the content library size in the cache of small-cell base stations. Furthermore, we show that range expansion, which is known to be of substantial value in wireless networks, is almost impotent to curb the caching inefficiency. Surprisingly, unlike the conventional wisdom recommending traffic offloading from macro cells to small cells, in cache-enabled HetNets, it is generally more beneficial to exclude offloading altogether or to do the opposite.**

## I. Introduction

Heterogeneous cellular networks (HetNets) are one of the key enablers for emerging cellular network systems to meet the exponential surge of wireless traffic demand [1], [2]. By substantially densifying the network, in particular in lower tiers, operators are empowered to sufficiently shrink the communication distances between base stations (BSs) and user equipments (UEs), enhancing the coverage performance and spectral efficiency. However, dense HetNets require expensive backhaul links between BSs and the core network (and/or among BSs), which may increase the total communication costs. Caching popular contents at the small-cell BSs has been suggested to reduce the reliance on backhaul [3]. Caching also improves the Quality-of-Service/Experience (QoS/QoE) for the UEs as the high-demand contents are placed near the UEs and thus accessible with a lower latency [4].

Content placement is, therefore, important and has been studied extensively for cellular networks. For a given topology of small-cells, the authors of [4] introduced the idea of substituting backhaul with caching in the BSs to reduce the network delay. They then developed an optimal femto-caching scheme for both uncoded and coded caching. Caching is also considered to improve energy-efficiency. Reference [5] optimized coded caching to minimize the energy consumption of backhaul and storage. Furthermore, [6] investigated the impact of caching on the energy-efficiency of video on-demand applications. Utilizing caching in fog radio access networks (F-RANs) was also shown in [7] to improve D2D communications efficiency in HetNets.

One way to measure caching efficiency is to evaluate the cache-hit probability (also referred to as *hit ratio* or hit rate). The hit ratio is defined as the probability that the requested content is successfully found/delivered from a cache, not the backhaul. The broadcast nature of wireless communications is exploited in [8] to introduce the optimal randomized caching in small-cell networks. It is shown in [8] that in many practical cases the hit ratio of randomized contents placement is much higher than that of the intuitive caching of the most popular contents everywhere. This is due mainly to the diversity of wireless medium and the fact that in HetNets each UE is likely to be located in the coverage area of multiple BSs [5]. The method in [8] is extended further in [9] to investigate the impact of content retransmission in small-cell networks on the hit ratio in both high and low mobility scenarios. For a given number of retransmission attempts, [8] then optimized the content placement to maximized the hit ratio.

Randomized content placement is extended further in [10], [11] to $K$-tier HetNets, where the probability of content placement stays the same across the BSs in each tier. The optimal probabilistic content placement is shown in [10], [11] to resemble a water-filling-type law.

Nevertheless, the above studies fail to incorporate the following two important practical aspects of the HetNets in caching performance analysis: (*i*) Although the cost of installing caching equipments (memory and the corresponding hardware) is much lower than that of the backhaul's for a dense/ultra-dense network, such as in 5G [1], the aggregated cost of caching may exceed the backhaul's cost; (*ii*) densification also amplifies the impact of excessive inter-cell interference as many UEs might receive interferences from a large number of BSs through a line-of-sight (LOS) channel [12], [13].

Analysis in [12] shows that by increasing the density of BSs, the coverage probability in cellular network reduces to zero, far lower than what the ideal standard-path-loss model [14] indicates. In such a case, regardless of efficiency of content placement, the UE cannot receive the content due to the low SIR. In such a case, the results of the current cellular network caching with a standard path-loss model, such as [6], [7], [8], [10], [11] are not directly applicable to dense HetNets.

We investigate the caching efficiency of dense HetNets to address the above two important aspects of HetNets. We adopt stochastic geometry as an analytical tool to investigate whether caching is beneficial in dense HetNets or not. To the best of our knowledge, this has not been investigated before. Our model incorporates the actual traits of signal propagation in modern cellular networks, i.e., LOS/NLOS path-loss model along with Nakagami fading for small-scale fading fluctuations. We also account for the costs of backhaul and caching, and analytically derive coverage probability, backhaul usage probability, hit-ratio, and area spectral efficiency (ASE), to provide quantitative insights on the impact of various system and design parameters.

Our analysis shows that while densification is beneficial in reducing the backhaul-usage probability and increasing ASE, it reduces cache-hit probability and increases the network cost. To provide a comprehensive performance evaluation of caching systems, we introduce caching efficiency as the ASE per cost. Our analysis indicates that in general caching is *not beneficial in densified networks*. We further observe that it is enough to cache only about $3\%$ of the global content library size in lower tiers. Furthermore, a common networking mechanism such as range expansion, which is shown to be of substantial value in conventional (no caching) networking via off-loading, could not alter this phenomenon. In a sharp contrast with the conventional HetNets in which off-loading to the small cells is suggested [1], [2], our analysis shows that in cache-enabled networks, only traffic offloading from the small cells to the micro cells improves the caching performance.

## II. System Model

### A. Network Model

We consider a dense HetNets with universal frequency reuse, complying with the interference-limited regime. Our focus is on the downlink of a cache-enabled $K$-tier HetNet, where $K$ tiers (classes/technologies) of BSs are randomly located in a 2-D plane [14], [10]. Tier $i$ is specified by a tuple, $(\lambda_i, P_i, \beta_i, S_i, \phi_i \in [0,1])$, where $\lambda_i$ is the BSs' spatial density, $P_i$ is their maximum transmission power, $\beta_i$ is the prescribed SIR threshold, $S_i$ is each BS's maximum cache storage, and $\phi_i \in [0,1]$ is an indicator of the adopted caching strategy (which will be elaborated further in Section IV).

In tier $i$, the spatial distribution of the BSs is modeled with a homogenous Poisson point process (HPPP), $\Phi_i \in \mathbb{R}^2$, with a spatial density of $\lambda_i \geq 0$, where $\Phi_i$ and $\Phi_j$ are mutually independent, for $\forall i, j, i \neq j$. In our model, the UEs are single-antenna and distributed according to a HPPP, $\Phi_U$, independent of $\Phi_i$, with a spatial density of $\lambda_U$. We further assume that $\lambda_U \geq \sum_i \lambda_i$, i.e., all the BSs are assumed to be active. Without loss of generality we investigate a *typical* UE, which is positioned at the origin and associated with BS $x_i$. This model can be easily extended to the users with multiple antennas as in [13].

### B. Caching Strategy

We consider a content library, $\mathcal{F} = \{f_1, f_2, \dots, f_F\}$ with the size of $F = |\mathcal{F}|$, where the files in this library are indexed based on their popularity, e.g., $f_c$ is the $c$-th most popular file. For simplicity, as in [8], [9], we also assume files of the same same size. BSs in tier $i$ are able to cache $S_i \leq F$ distinctive files.

A file can be either cached based on their popularity, $s_i = \mathrm{pop}$, or randomly $s_i = \mathrm{rnd}$. In the former case, the first $S_i$ most popular contents (MPCs) are cached. In the latter case, or random content selection (RCS), each BS randomly draws an index $c \in [1, F - S_i + 1]$, with probability $\frac{1}{F - S_i + 1}$ and caches contents with indices in $[c, c + S_i]$. We further assume that BSs of tier $i$ randomly and independently choose their caching strategies $s_i \in \{\mathrm{pop}, \mathrm{rnd}\}$, where $\phi_i = \mathbb{P}\{s_i = \mathrm{pop}\} \in [0,1]$.

One way to evaluate the efficiency of the caching strategy is the *hit ratio*, or the probability that the requested content is available in the cache and successfully delivered. Note that RCS may seem counterproductive since one expects the contents with higher popularity to be requested more often. It is, however, shown in [8], [11] that in order to maximize the hit ratio, it is not necessarily optimal to adopt the MPC scheme, particularly in HetNets that the typical UE is likely to be located in the coverage of several adjacent BSs.

The set of BSs in tier $i$ which cache $f_c$, $\Phi_i[c]$, is also a HPPP with density $q_i[c]\lambda_i$, where $q_i[c]$ is the probability that $f_c$ is cached at each BS in tier $i$:

$$q_i[c] = \mathbb{P}\{c \in \mathcal{S}_{x_i}\} = \phi_i 1_{c \leq S_i} + (1 - \phi_i) \sum_{m=1}^{F - S_i + 1} \frac{1_{m \leq c \leq m + S_i}}{F - S_i + 1}$$

$$= \begin{cases} \phi_i + \frac{(1-\phi_i)c}{F - S_i + 1}, & 1 \leq c \leq S_i, \\ \frac{(1-\phi_i)S_i}{F - S_i + 1}, & S_i < c \leq F - S_i + 1, \\ \frac{(1-\phi_i)}{F - S_i + 1}(F - c + 1), & c > F - S_i + 1. \end{cases} \quad (1)$$

The content popularity is characterized with a Zipf distribution as in [9]. So, the probability of $f_c$ being requested, $a_c$ is

$$a_c = \frac{c^{-\kappa}}{\sum_{n=1}^{F} n^{-\kappa}},$$

where $\kappa \geq 0$ is the shape parameter of the distribution, also referred to as the *popularity exponent*. For $\kappa \to 0$, the content popularity reduces to the uniform distribution. For a large $\kappa$, however, the most popular contents have much higher chance to be requested.

### C. Channel Model

We consider a narrow-band, block-fading channel in which fading evolves randomly according to a specified fading distribution at the start of each frame and remains unchanged throughout the frame transmission. The channel model comprises of a large-scale path-loss and a small-scale fading component. The received signal at the typical UE originated from BS $x_i$ undergoes LOS or NLOS path-loss attenuation, depending on its relative distance to the UE, density of buildings, etc. To model the path-loss environment, we adopt the 3GPP path-loss model [15], [13]:

$$L_i(\|x_i\|) = \begin{cases} L_i^L(\|x_i\|) = \frac{\phi_L}{(1+\|x_i\|)^{\alpha_i^L}}, \sim p_i^L(\|x_i\|), \\ L_i^N(\|x_i\|) = \frac{\phi_N}{(1+\|x_i\|)^{\alpha_i^N}}, \sim p_i^N(\|x_i\|), \end{cases} \tag{2}$$

where $p_i^N(\|x_i\|) = 1 - p_i^L(\|x_i\|)$ is the probability that the link between BS $x_i$ and the typical UE is in NLOS mode. Here, we assume that LOS probabilities are independent across BSs. We consider the ITU-R UMi model in [15], where the LOS probability is specified as

$$p_i^L(\|x_i\|) = \min\left\{\frac{D_0^i}{\|x_i\|}, 1\right\}\left(1 - e^{-\frac{\|x_i\|}{D_1^i}}\right) + e^{-\frac{\|x_i\|}{D_1^i}}, \tag{3}$$

and $D_0^i$ and $D_1^i$ characterize the near-field (LOS) and far-field (NLOS) critical distances, respectively. Therefore, if $\|x_i\| \leq D_0^i$, then BS $x_i$ is in LOS mode. For $\|x_i\| > D_0^i$, the probability of LOS mode declines exponentially with the distance, and for $\|x_i\| \gg D_1^i$, it converges to 0.

In (2), for $n_i \in \{L, N\}$, $\alpha_i^L$ (resp. $\alpha_i^N$) is the path-loss exponent associated with the LOS (resp. NLOS) link where $2 < \alpha_i^L < \alpha_i^N \leq 8$, $\phi_i^L$ (resp. $\phi_i^N$) is a constant, characterizing the LOS (resp. NLOS) wireless propagation environment, and is related to various factors, such as the height of transceivers, antenna's beam-width, weather, etc. Small-scale fading is modeled using normalized Nakagami fading:

$$H_{x_i} = \begin{cases} H_{x_i}^L = \Gamma(M_i^L, \frac{1}{M_i^L}), \sim p_i^L(\|x_i\|), \\ H_{x_i}^N = \Gamma(M_i^N, \frac{1}{M_i^N}), \sim p_i^N(\|x_i\|), \end{cases} \tag{4}$$

where $\Gamma(a, 1/a)$ is normalized Nakagami distribution with parameter $a$. Depending on whether the link is LOS or NLOS, different parameters are considered for the Nakagami fading. In general, we expect $M_i^L > M_i^N$, as the LOS links often fluctuate less severely.

### D. Simulation Model and Parameters

We adopt the Monte Carlo technique for the simulation and numerical study. We consider a 2-tier HetNet, $K = 2$, where the transmit power of the macro BSs in the first and second tiers are $P_1 = 40$W, and $P_2 = 4$W, respectively. The LOS (resp. NLOS) path-loss exponent is $\alpha_1^L = \alpha_2^L = 2.4$ ($\alpha_1^N = \alpha_2^N$=4). The path-loss intercept parameters are set to 1. Also, we set $D_1^0 = 80$m, $D_1^1 = 164$m, $D_2^0 = 16$m, and $D_2^1 = 36$m. The SIR thresholds are $\beta_1 = 2$ and $\beta_1 = 4$.

The size of content library is set to $F = 100$ fixed-length files, and cache sizes are fixed at $S_1 = 20$ and $S_2 = 5$ fixed-length files. The BSs in each tier are randomly distributed within a disk with radius $10,000$ units according to the corresponding tier densities, where $\lambda_1 = 10^{-3}$ BSs per square kilometers. The presented results are based on analysis of $40,000$ simulation snapshots. The other parameters not specified above are either design parameters or defined for each particular experiment.

## III. Content-Aware Max-SIR Cell Association

Suppose the typical UE requests content $f_c$, the signal-to-interference ratio (SIR) experienced at the typical UE served by BS $x_i \in \Phi_i[c]$ is

$$\mathrm{SIR}_{x_i}[c] = \frac{P_i L_i(\|x_i\|) H_{x_i}}{\sum_{j=1}^K I_j}, \tag{5}$$

where the interference of tier $j$, $I_j$, is a shot noise process,

$$I_j = \sum_{x_j \in \Phi_j[c] \setminus x_0} P_j L_j(\|x_j\|) H_{x_j} + \sum_{x_j \in \Phi_j \setminus \Phi_j[c]} P_j L_j(\|x_j\|) H_{x_j}$$

$$= \sum_{x_j \in \Phi_j \setminus x_0} P_j L_j(\|x_j\|) H_{x_j}, \tag{6}$$

which is independent of the requested file. The typical UE successfully receives the data transmitted by BS $x_i$, if the corresponding SIR is larger than the SIR threshold, $\beta_i > 0$. The coverage probability is then equal to the complementary cumulative distribution function (CCDF) of the SIR.

The UE requesting $f_c$ should be associated with a cell with $f_c$ cached in its corresponding BSs. Such an association can be made based on different criteria. We consider Max-SIR cell association (CA) which is shown to provide the maximum coverage performance, see, e.g., [14], [16], [13]. Without considering the availability of $f_c$, Max-SIR CA associates the typical UE with the BS that provides the highest SIR, regardless of whether $f_c$ is cached, or retrieved via the backhaul. To extend Max-SIR CA incorporating the availability of the content, $f_c$, we define

$$\mathcal{A}_c = \left\{\exists i : \max_{x_i \in \Phi_i[c], \forall i} \mathrm{SIR}_{x_i}[c] \geq \beta_i\right\}, \tag{7}$$

as the set of BSs with $f_c$ in their cache providing acceptable level of SIR for the UE. There is a BS in the network to be associated with the UE if $\mathcal{A}_c \neq \varnothing$. Content-aware cell association is expected to be effective in reducing the backhaul usage [3].

Using the same line of argument as in [14], the corresponding coverage probability, $\varrho_c$, is upper-bounded as

$$\varrho_c = \mathbb{P}\left\{\max_{x_i \in \bigcup_{i=1}^K \Phi_i[c]} \mathrm{SIR}_{x_i}[c] \geq \beta_i\right\}$$

$$\leq \sum_{i=1}^K \mathbb{E} \sum_{x_i \in \Phi_i[c]} 1\left(\mathrm{SIR}_{x_i}[c] \geq \beta_i\right) = 2\pi \sum_{i=1}^K q_c[i] \varrho_i[c], \tag{8}$$

where $\varrho_i[c] = \lambda_i \int_0^\infty r_i \mathbb{P}\left\{\mathrm{SIR}_{x_i}[c] \geq \beta_i\right\} \mathrm{d}r_i$ and the equality holds for $\beta_i \geq 1$. Using (5), we write

$$\frac{\varrho_i[c]}{\lambda_i} = \int_0^\infty x_i \mathbb{P}\left\{\frac{P_i L_i(x_i) H_{x_i}}{\sum_{j=1}^K I_j} \geq \beta_i\right\} dx_i$$

$$= \sum_{n_i \in \{L,N\}} \int_0^\infty x_i p_i^{n_i}(x_i) \mathbb{P}\left\{\frac{L_i^{n_i}(x_i) H_{x_i}^{n_i}}{\sum_j I_j} \geq \frac{\beta_i}{P_i}\right\} dx_i$$

$$\leq \sum_{n_i\in\{L,N\}} \int_0^\infty x_i p_i^{n_i}(x_i)\mathbb{E}\left(1-\left(1-e^{-\frac{v_i M_i^{n_i}\beta_i}{P_i L_i^{n_i}(x_i)}\sum_j I_j}\right)^{M_i^{n_i}}\right)dx_i$$

$$= \sum_{n_i\in\{L,N\}} \sum_{m_i=1}^{M_i^{n_i}} \binom{M_i^{n_i}}{m_i}(-1)^{m_i+1}\int_0^\infty p_i^{n_i}(x_i)\prod_{j=1}^{K}$$

$$\times\mathbb{E}_{I_j} e^{-\frac{\beta_i m_i v_i M_i^{n_i}}{P_i L_i^{n_i}(x_i)} I_j} dx_i, \tag{9}$$

where the inequality is due to Alzer's Lemma [17] and $v_i = M_i^{n_i}(M_i^{n_i}!)^{-1/M_i^{n_i}}$.

Noting that the fading is normalized Nakagami, it is straightforward to show

$$\mathbb{E}_{I_j}[e^{-tI_j}] = \left(\mathbb{E}e^{-t\sum_{x_j\in\Phi_j\backslash x} P_j L_j(\|x_j\|)H_{x_j}}\right)$$

$$= \mathbb{E}_{\Phi_j}\prod_{x_j\in\Phi_j}\left(\sum_{n_j\in\{L,N\}}\frac{p_j^{n_j}(\|x_j\|)}{(1+\frac{tP_jL_j^{n_j}(\|x_j\|)}{M_j^{n_j}})^{M_j^{n_j}}}\right)$$

$$= e^{-2\pi\lambda_j \sum_{n_j\in\{L,N\}}\int_0^\infty y_j p_j^{n_j}(y_j)\left(1-(1+\frac{tP_jL_j^{n_j}(y_j)}{M_j^{n_j}})^{-M_j^{n_j}}\right)dy_j},$$

where in the first step, we note that LOS/NLOS modes are independent across the BSs and the fading power gains are i.i.d. In the next step we use the Laplace generation functional of HPPP [17]. Substituting the above into (9) and setting $t = \frac{\beta_i m_i v_i M_i^{n_i}}{P_i L_i^{n_i}(x_i)}$,

$$\varrho_i[c] \leq \lambda_i \sum_{n_i\in\{L,N\}} \sum_{m_i=1}^{M_i^{n_i}} \binom{M_i^{n_i}}{m_i}(-1)^{m_i+1}\int_0^\infty p_i^{n_i}(x_i) \tag{10}$$

$$\times e^{-2\pi\lambda_j\sum_{n_j\in\{L,N\}}\int_0^\infty y_j p_j^{n_j}(y_j)\left(1-(1+\frac{v_i m_i M_i^{n_i}\beta_i P_j L_j^{n_j}(y_j)}{P_i L_i^{n_i}(x_i) M_j^{n_j}})^{-M_j^{n_j}}\right)dy_j} dx_i.$$

Inserting (10) into (8) and obtaining the summation over the content request probability, we obtain the coverage probability as

$$\varrho = \sum_{c=1}^{F} a_c \sum_{i=1}^{K} q_i[c]\varrho_i[c], \tag{11}$$

which is a function of system parameters including the density of the BSs, the library and cache size, popularity exponent, and the SIR thresholds.

Fig. 1 shows the accuracy of the derived upper-bound while comparing our simulation results with the coverage probability in (11). The simulation parameters are given in Section II.D.

In practical cases for high-capacity HetNets, where $\beta_2 \geq 1$ [14], the upper-bound becomes very tight. Even for $\beta_2 = 0.5$, the upper-bound closely follows the simulation. Furthermore, as shown in both plots, increasing $\beta_2$ reduces the coverage probability. Fig. 1-a also indicates that by increasing the LOS path-loss exponent, the coverage probability is slightly reduced. Fig. 1-b, shows that for $\beta_2 \lessapprox 3$, densification improves the coverage performance, whereas for $\beta_2 \gtrapprox 3$, increasing $\lambda_2$ reduces the coverage probability, making densification detrimental to the coverage performance. This phenomenon has been explored extensively in the literature of HetNets, see, e.g., [12], [13], and has been attributed to the LOS component of interfering signals.

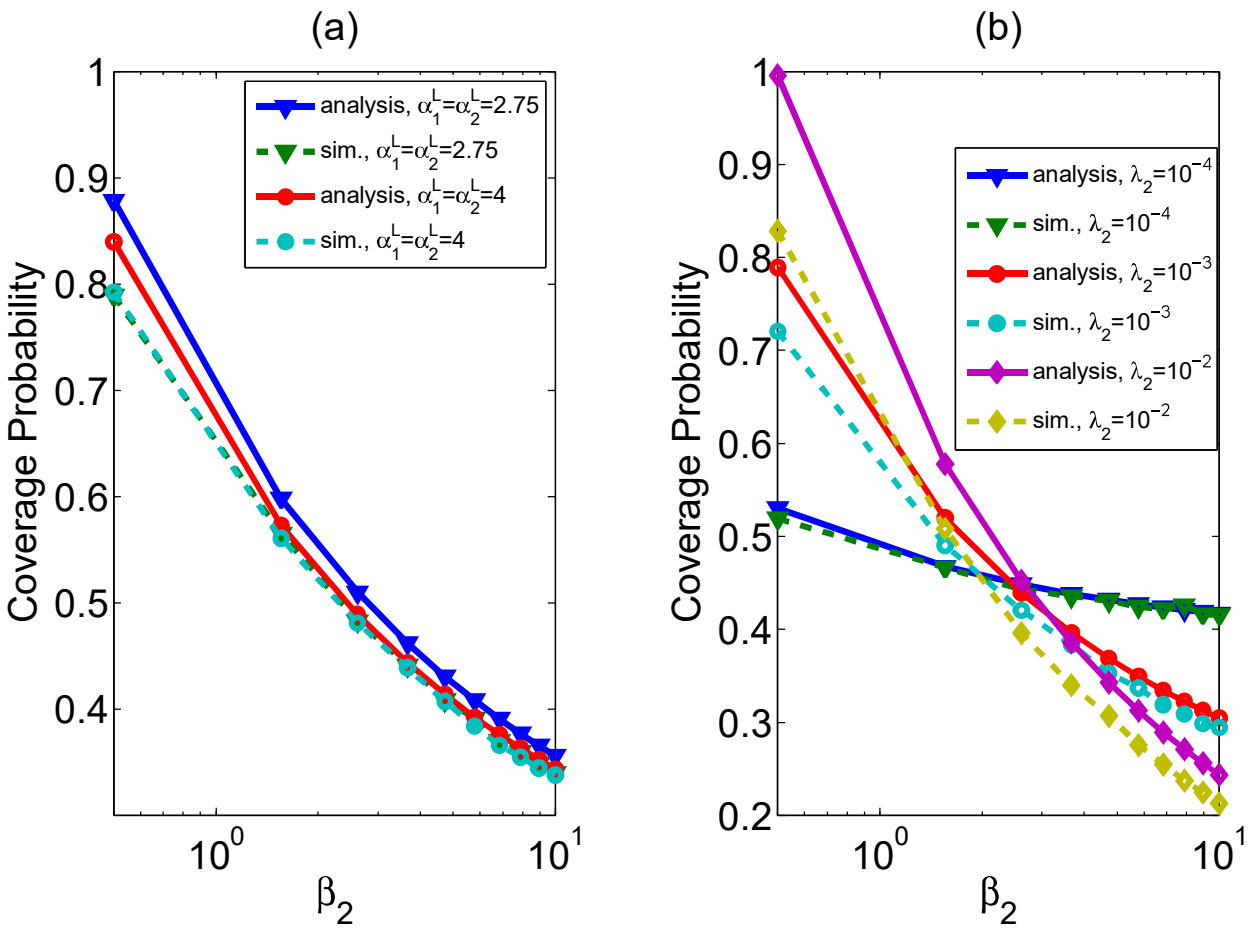

Fig. 1. Coverage probability versus the SIR threshold of tier 2 $\beta_2$. Simulation parameters are given in Section II.D.

In fact, for the dense networks, there is always an unvanishing interference which is at least as large as the attending signal. Therefore, in some cases, regardless of the distance between the associated BS and the typical UE, the SIR could not improve further. In what follows, we show that this phenomenon remains harmful in cache-enabled HetNets. This has not been discussed before in the related literature.

## IV. Caching Performance

### A. Cache Hit vs. Backhaul Usage

Caching improves the efficiency of content delivery by reducing the backhaul usage. One way to assess the efficiency of a caching system is to evaluate the cache-hit probability (or hit ratio/rate) and backhaul-usage probability. The cache-hit probability is defined as the probability that the required content is found in the cache of a BS and successfully delivered. The backhaul-usage probability is the probability that the requested data is obtained from the core network via the backhaul. In cache-enabled HetNets, a reasonable design objective is to minimize the latter and maximize the former [3]. We dissect the coverage probability in (11) as

$$\sum_i \varrho_i[c] = \sum_{i=1}^{K} q_i[c]\varrho_i[c] + (1-q_1[c])\varrho_1[c],$$

in which $p_{\mathrm{hit}}[c] = \sum_i\sum_c a_c q_i[c]\varrho_i[c]$ is the cache-hit probability, and $p_{\mathrm{bh}}[c] = \sum_c a_c(1-q_1[c])\varrho_1[c]$ is the backhaul-usage probability. Therefore,

$$\sum_i \varrho_i[c] = p_{\mathrm{hit}}[c] + p_{\mathrm{bh}}[c],$$

as $f_c$ is either cached or retrieved from the core via the backhaul.

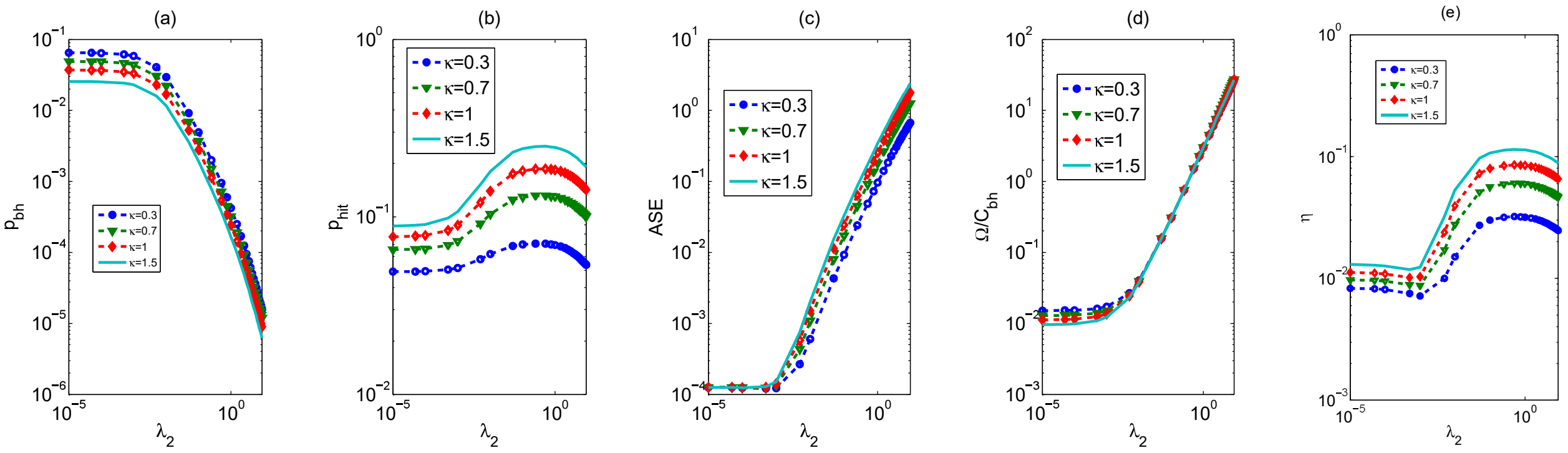

Fig. 2. (a) Backhaul-usage probability, $p_{\rm bh}$, versus Tier 2 BSs' density, $\lambda_2$; (b) Caching-hit probability, $p_{\rm hit}$, versus $\lambda_2$; (c) ASE versus $\lambda_2$; (d) Network cost, $\Omega$, versus $\lambda_2$; (e) Caching efficiency, $\eta$, versus $\lambda_2$.

Fig. (2)-a plots the backhaul-usage probability versus $\lambda_2$ for different values of the popularity exponent, $\kappa$. The backhaul-usage probability is shown to be improved (reduced) by increasing the popularity exponent. Also, densification in tier 2 is shown to reduce the backhaul-usage probability significantly.

We further look at the caching performance from the hit-ratio perspective. As shown in Fig. (2)-b, densification does not consistently improve the hit ratio. In fact, for a sparse ($\lambda_2 < 10^{-3}$) to a moderately dense ($10^{-3} < \lambda_2 < 1$) tier 2, the hit ratio improves as the UEs expect to receive their requested content *successfully* from the cache. Here, the typical UE can often find the best BS (in terms of SIR) that also has the requested content. For $\lambda_2 > 1$—dense configuration—increasing $\lambda_2$, however, reduces the hit ratio due mainly to excessive LOS interference. In this case, regardless of how close the contents are located to the UEs, or how efficient the contents are placed, the negative impact of the interference dominates the hit ratio. This shows that many previous studies of cache-enabled systems (e.g., [4], [8], [10]) are only applicable to moderately dense networks, where standard path-loss model is still valid.

### B. Area Spectral Efficiency

Another crucial performance metric in HetNets is ASE [14], [17]. ASE measures the average aggregate data rate provided per unit area (bps/Hz/m$^2$):

$$\overline{R} = \sum_{c=1}^{F} a_c \left( \sum_{i=1}^{K} q_i[c] \varrho_i[c] \lambda_i R_i + (1 - q_1[c]) \varrho_1[c] \lambda_1 R_1 \right). \tag{12}$$

In the inner summation of (12), the first and second terms are attributed to the caching and backhaul performance, respectively.

Fig. (2)-c plots ASE vs. $\lambda_2$. Densification is shown to substantially increase ASE (almost linearly). Further, ASE is increased by increasing the popularity exponent.

### C. Cost Per Unit Area

In a cache-enabled dense HetNet, content delivery needs to be planned carefully to keep the costs at an acceptable level. Backhaul connectivity is often provided through a network of optical fibers. In urban area, however, deployment and maintenance costs of such networks are very high. Furthermore, there is an extra cost associated of caching. In what follows, we formulate the cost per unit area.

Let $C_{\rm bh}$, and $C_{\rm s}$ denote the generic costs of the backhaul and caching, respectively, including installation, maintenance, operational costs, etc. Due to the nature of the technology, it is reasonable to assume that $C_{\rm s} \ll C_{\rm bh}$. The aggregated caching cost per unit area of coverage in a dense HetNet, $\Omega$, is

$$\Omega = \lambda_1 (F - S_1) C_{\rm bh} \sum_{c=1}^{F} a_c p_{\rm bh}[c] + C_{\rm s} \sum_{i=1}^{K} \lambda_i S_i, \tag{13}$$

where the first term is the cost of using backhaul which depends on the backhaul usage (represented by the backhaul-usage probability), and the second term is the aggregated cost of caching (represented by the caching storage capacity).

We investigate the impact of densification on the cost of coverage per unit area. Fig. (2)-d plots $\Omega/C_{bh}$ vs. $\lambda_2$, where we assume $C_{\rm s} = 0.01 C_{\rm bh}$. By increasing $\lambda_2$ (densification of tier 2), the cost is shown to monotonically increase despite the fact that the backhaul-usage probability, $p_{\rm bh}$, becomes considerably smaller (Fig. (2)-a). In this case, although the usage of backhaul is reduced with an effective caching strategy, the cost kept on increasing because in dense networks, the accumulative cost of caching eventually dominates the backhaul cost. From (2)-b one can also see that for such a high cost, the hit ratio is also low. Therefore, caching in a densified HetNet only cannot be considered as a solution for the high cost of the backhaul.

## V. Caching in Dense HetNets

As shown in Section IV, densification may often have negative impact on the performance of the dense HetNets, in terms of coverage probability, hit rate, and the network cost. So, to provide a clear picture of the impact of caching in the dense HetNets, we incorporate the above performance metrics in defining a new caching efficiency measure, $\eta$, which

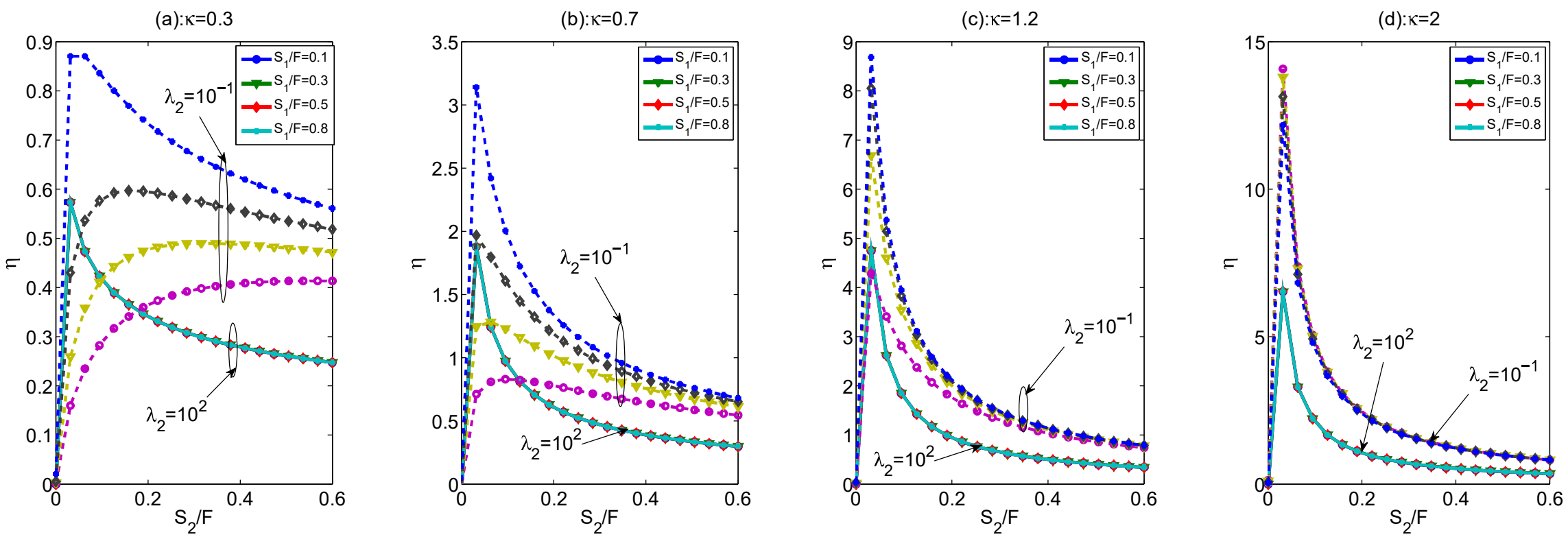

Fig. 4. Impact of cache size on $\eta$.

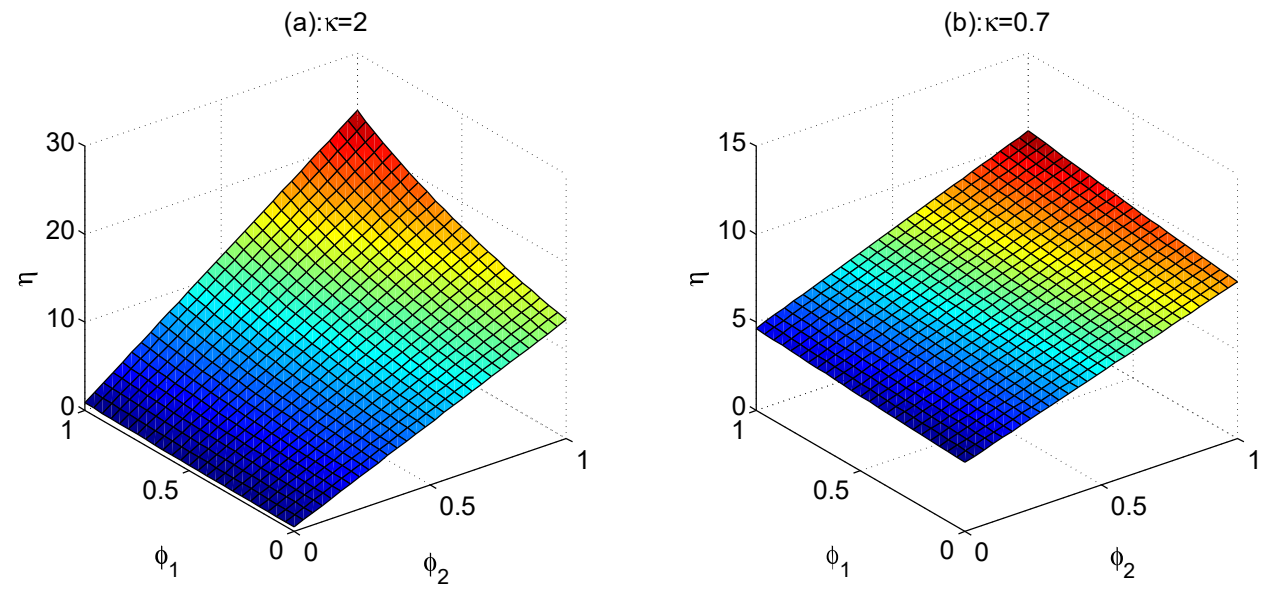

Fig. 3. Impact of probabilities $\phi_1$ and $\phi_2$ on caching efficiency, $\eta$.

indicates the ASE per cost:

$$\eta = \frac{\overline{R}}{\Omega}, \tag{14}$$

where $\overline{R}$ is the ASE as in (12), and $\Omega$ is the cost per unit area as in (13). An ideal design is to maximize the spectral efficiency while lowering the costs, i.e., maximizing $\eta$.

*1) **Caching is Beneficial in Moderately Dense HetNets**:* Fig. (2)-e plots $\eta$ vs. $\lambda_2$ for several values of content popularity exponent, $\kappa$. For $\lambda_2 > 1$, densification is shown to reduce the caching efficiency due mainly to the high cost of caching and weak coverage performance. Nevertheless, moderate densification of a sparse network (from $\lambda_2 < 10^{-3}$ to $10^{-3} < \lambda_2 < 1$) improves the caching efficiency. In such a case, the high ASE compensates for the negative impact of the high caching cost and low hit ratio.

*2) **Caching Contents across Tiers**:* The best caching strategy is obtained via the following optimization:

$$\max_{\phi_i \in [0,1] \forall i} \eta.$$

Note that if $\phi_i = 1$, then BSs of tier $i$ only cache the most popular contents, while $\phi_i = 0$ means the BSs cache randomly. Fig. 3 plots $\eta$ for different content popularity exponent and $\phi_i$ while Fig. 3 shows that MPC always outperforms RCS, suggesting use of MPC across all tiers.

*3) **Impact of Cache Size**:* Fig. 4 plots $\eta$ vs. $\overline{S}_2 \triangleq \frac{S_2}{F} \leq 1$ for several values of $\overline{S}_1 \triangleq \frac{S_1}{F} \leq 1$. For cases of moderately densified HetNets , i.e., $\lambda_2 = 10^{-1}$, there is an optimal caching size in tier 2 that maximizes the caching efficiency. For a $\kappa \geq 0.7$, the optimal cache size is fairly small compared to the library size. In fact, Fig. 4-a shows that the optimal cache size is only $3\%$ of the most popular contents.

Figs. 4-b-d also indicate that the optimal cache size is independent of the popularity exponent $\kappa$. For a small $\kappa$, Fig. 4-a further suggests that increasing the cache size in tier 2 improves $\eta$. However, one can afford increasing the cache size in tier 1, and the cache size of tier 2 can then be reduced to $3\%$ of the size of the content library.

We further oberve that the cache size in tier 1 has a substantial impact on the caching efficiency, especially for $\kappa \leq 1.2$, where increasing $S_1$ to up to $80\%$ of the content library is shown to improve the caching efficiency considerably. For a larger $\kappa$, however, (see Fig. 4-d), it is sufficient to merely cache $10\%$ of most popular contents in tier 1. In either case, Fig. 4 suggests that for a given $\kappa$, caching efficiency is improved by carefully selecting the caching size.

In a dense HetNet where $\lambda_2 = 10^2$, the optimal cache size in tier 2 is almost $3\%$ of the size of the content library, regardless of the parameter $\kappa$. The caching efficiency is not related to the cache size in tier 1 either. This is in sharp contrast with the case of moderately dense HetNets, while the caching performance is also substantially lower than that of moderately dense HetNets.

*4) **Impact of Traffic Offloading**:* In our analysis, we adopted content-aware Max-SIR CA rule as it is shown to be effective in reducing the backhaul usage as one of the main objectives of caching [3]. The above results, however, suggest that caching is not beneficial in dense HetNets. Therefore, there seems to be a gap in the literature on how to enhance the caching efficiency in dense HetNets.

We investigate this important issue by introducing range expansion into our analysis. We consider the range expansion parameters, $\rho_i \in (0, 1]$, where $\sum_i \rho_i = 1$. We then substitute

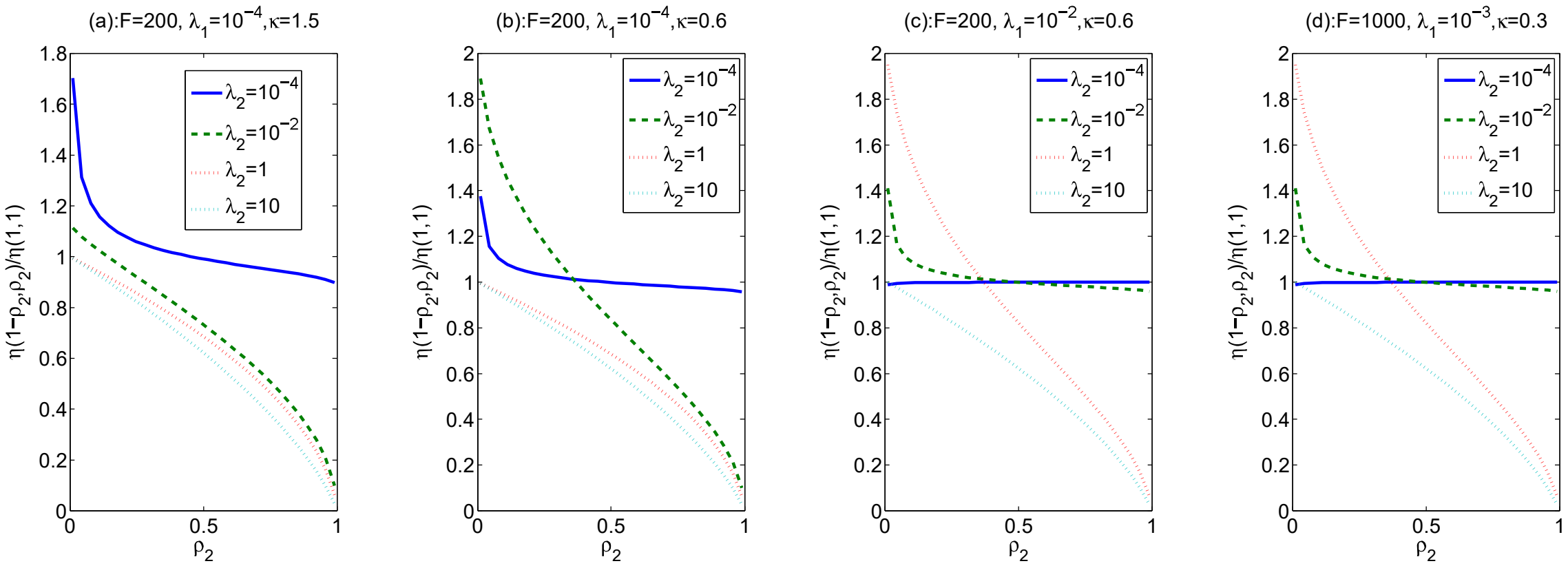

Fig. 5. Impact of range expansion parameter $\rho_2$ ($\rho_1 = 1 - \rho_2$) on $\eta$.

the SIR thresholds, $\{\beta_i\}$, with their scaled versions, $\frac{\beta_i}{\rho_i} \geq \beta_i$. Using a smaller value of $\rho_i$, a user is less likely to be associated with tier $i$. Note that using this modified version of the CA, for the typical UE associated with a tier $j$ BS, the data transmission rate is not affected and is equal to $\log(1+\beta_j)$

To investigate the impact of range expansion on the caching efficiency we denote $\eta(\rho_1, \rho_2, \ldots, \rho_K)$ as the caching efficiency for given set of range expansion parameters $(\rho_1, \rho_2, \ldots, \rho_K)$. In our formulation, $\eta(1, 1, \ldots, 1)$ is the caching efficiency under the content-aware Max-SIR CA policy as in (7).

To study the impact of range expansion on the caching efficiency, Fig. 5 shows $\frac{\eta(1-\rho_2,\rho_2)}{\eta(1,1)}$ vs. $\rho_2$, for $K = 2$, where $C_{\rm s} = .001 C_{\rm bh}$. Fig. 5 shows that for a sparse to moderately dense HetNet, i.e., $\lambda_2 < 1$, one can choose a $\rho_2$ that improves the caching efficiency. To improve the caching efficiency, Fig. 5 counter-intuitively suggests reduction of $\rho_2$, i.e., offloading traffic from small cells to the macro cells.

Our simulation results also show that offloading improves the caching efficiency if tier 2 is mildly densified, i.e., $10^{-3} < \lambda_2 < 1$, see, Figs. 5-c—5-d. Fig. 5 further indicates that for dense HetNets, i.e., $\lambda_2 > 1$, range expansion does not improve the caching efficiency.

## VI. Conclusions

In this paper we have studied the caching performance in dense HetNets. Our analysis incorporated the actual traits of dense cellular networks including the LOS/NLOS path-loss model, backhaul and caching costs, and provided performance metrics such as coverage probability, backhaul-usage probability, caching-hit probability, ASE, and the network cost. Our analysis showed that while densification is beneficial in reducing the backhaul-usage probability and increasing ASE, it reduces cache-hit probability and increases the network cost. To provide a comprehensive performance evaluation of caching systems, we then introduced caching efficiency by incorporating the above-mentioned performance metrics. Our analysis showed that caching is, in general, *not beneficial in densified networks*. We further observed that it is enough to cache only about 3% of the library size in tier 2. Furthermore, a common networking mechanism such as range expansion, which is shown to be of substantial value in conventional (no caching) networking, could not alter this phenomenon. We also showed that in sharp contrast with the conventional networks, in cache-enabled HetNets one should offload the traffic from small cells to macro cells. Our results also suggest that enabling caching benefits in dense HetNets needs further investigation of the impact of content placement as well as interference management.